\documentclass{ctr_summer}

\usepackage{ctrfont}
\usepackage{natbib}
\usepackage{undertilde}
\usepackage{color}

\usepackage{graphicx}

\usepackage{psfrag}
\usepackage{epsfig}
\usepackage{amsmath,rotating}
\usepackage{dashrule}
\usepackage{undertilde}

\usepackage{url}

\usepackage{color}

\usepackage{mhchem}

\newcommand{\diff}{\mathrm{d}}
\newcommand{\f}{\frac}
\newcommand{\ov}{\overline}
\newcommand{\ii}{\mathrm{i}}

\newcommand{\gtrsim}{\raisebox{-0.13cm}{~\shortstack{$>$ \\[-0.07cm]
      $\sim$}}~}

\makeatletter
\newcommand\footnoteref[1]{\protected@xdef\@thefnmark{\ref{#1}}\@footnotemark}
\makeatother

\title{Spectral analysis of multidimensional current-driven plasma instabilities and turbulence in hollow cathode plumes}

\shorttitle{Multidimensional current-driven plasma instabilities and turbulence}

\author{W. H. R. Chan\footnote[1]{\label{foot:cu}Department of Aerospace Engineering Sciences, University of Colorado Boulder}, K. Hara, J. M. Wang, S. S. Jain, S. Mirjalili \and I. D. Boyd\footnoteref{foot:cu}}

\shortauthor{Chan et~al.}

\begin{document}

\setcounter{page}{1}

\maketitle

Large-amplitude current-driven instabilities in hollow cathode plumes can generate energetic ions responsible for cathode sputtering and spacecraft degradation. A 2D2V (two dimensions each in configuration [D] and velocity [V] spaces) grid-based Vlasov--Poisson (direct kinetic) solver is used to study their growth and saturation, which comprises four stages: linear growth, quasilinear resonance, nonlinear fill-in, and saturated turbulence. The linear modal growth rate, nonlinear saturation process, and ion velocity and energy distribution features in the turbulent regime are analyzed. Backstreaming ions are generated for large electron drifts, several ion acoustic periods after the potential field becomes turbulent. Interscale phase-space transfer and locality are analyzed for the Vlasov equation. The multidimensional study sheds light on the interactions between longitudinal and transverse plasma instabilities, as well as the inception of plasma turbulence.\\

\hrule

\section{Introduction}\label{sec:intro}
  
Hollow cathodes are crucial for the production of plasmas, and particularly electrons, in electric spacecraft thrusters. The erosion of cathode structures can limit the lifetime of outerspace missions to less than $O(10{,}000)$ hours~\citep[e.g.,][]{friedly1992high, kameyama2000measurements, williams2000laser, mikellides2005hollow, mikellides2007evidence, mikellides2008wearII, goebel2007potential, jorns2014ion, lev2019recent}. A key cause of such sputtering is the generation of fast ions at the cathode orifice, with energies corresponding to up to $100\text{ eV}$~\citep{friedly1992high, williams1992electron, kameyama2000measurements, williams2000laser, boyd2004modeling, goebel2007potential, mikellides2008wearII, farnell2011comparison}. These high-energy ions are currently postulated to arise in part from collisionless and electrostatic current-carrying instabilities~\citep{williams2000laser, mikellides2005hollow, mikellides2007evidence, mikellides2008wearII, goebel2007potential, jorns2014ion, lopezortega2016importance, jorns2017propagation, sary2017hollowI, sary2017hollowII, hara2018test, lopezortega2018hollow, hara2019overview}. The generation of axially energetic ions has been numerically investigated through grid-based Vlasov--Poisson (direct kinetic) simulations in a single spatial dimension (1D)~\citep{hara2019ion, vazsonyi2020non}. We extend the analysis to two spatial dimensions (2D) to probe transversely (radially) energetic ions, whose presence has been observed experimentally~\citep{boyd2004modeling, goebel2007potential, farnell2011comparison, hall2019effect}. Such ions are already deflected from the centerline from their inception and can impinge spacecraft more easily.

\setcitestyle{notesep={ }}
Two categories of current-carrying instabilities in fully ionized plasmas are typically considered~\citep{omura2003particle, mikellides2005hollow}. The ion acoustic instability manifests when the electron drift speed $U_e$ exceeds the Bohm speed $\sqrt{k_\text{B} T_e/m_i}$ and $T_e \gg T_i$, where $k_\text{B}$, $T_e$, $T_i$, and $m_i$ are the Boltzmann constant, electron temperature, ion temperature, and ion mass, respectively~\citep[and references therein]{stringer1964electrostatic}. Physically, ion oscillations are excited by the net electron drift. Over a broad range of $T_i/T_e$, the Buneman instability arises when $U_e \gtrsim c_e = \sqrt{k_\text{B} T_e/m_e}$, where $c_e$ and $m_e$ are the electron thermal speed and electron mass, respectively~\citep{buneman1959dissipation}. For 1D, the threshold is about $U_e \geq 1.3 c_e$. Physically, electron oscillations are excited by the net electron drift. We consider $T_e/T_i = 10$, which is representative of cathode operating conditions~\citep{goebel2005hollow, mikellides2005hollow, mikellides2007evidence, mikellides2008wearII, farnell2011comparison, jorns2017propagation}. Here, both instabilities can be physically relevant depending on $U_e$. We build on previous studies of the two-dimensional Buneman instability~\citep{amano2009nonlinear} but focus on ion acceleration instead of electron acceleration and use a physical mass ratio corresponding to a hydrogen plasma ($m_i/m_e = 1.8229\times10^3$).
\setcitestyle{notesep={, }}

The objective of this work is to analyze the stages of instability growth, as well as the inception of multidimensional plasma turbulence and its spectral characteristics. In Section~\ref{sec:setup}, we describe our computational and physical setup. In Section~\ref{sec:1D}, we revisit key results from the simpler 1D instability to obtain physical insights. These are used to interpret results of the 2D instability in Section~\ref{sec:2D}. Conclusions are provided in Section~\ref{sec:conc}.

%
%
%
%

\section{Methodology}\label{sec:setup}

\subsection{Direct kinetic solver}

\setcitestyle{notesep={; }}
The direct kinetic solver employed here was originally developed at the University of Michigan with verification and validation against canonical and complex plasma problems, such as waves, electron-emitting sheaths, and Hall thruster discharges~\citep[and references therein]{hara2018test, raisanen2019two, vazsonyi2020non}. In contrast to state-of-the-art particle-in-cell solvers, direct kinetic solvers eliminate statistical noise and are suitable for investigating instability growth and turbulence inception. Under the electrostatic approximation, the solver computes the time evolution of the probability density function, $f_*$, for some particle type~$*=i,e$ according to the following transport equation
\begin{equation}
\f{\partial f_* (\mathbf{x},\mathbf{v};t)}{\partial t} + \mathbf{v}\cdot\nabla_\mathbf{x} f_*(\mathbf{x},\mathbf{v};t) + \f{q_* \mathbf{E}}{m_*}\cdot\nabla_\mathbf{v} f_*(\mathbf{x},\mathbf{v};t) = 0,
\end{equation}
where $\mathbf{x}$, $\mathbf{v}$, and $\mathbf{E}$ respectively denote the position, velocity, and electric field vectors, $q_*$ denotes the charge of the simulated particle type, and $t$ denotes the time. The computational domain is discretized in $\mathbf{x}$--$\mathbf{v}$ space with a parallelized second-order finite-volume method, which is described by~\citet{chan2022grid,chan2022enabling}. Gauss's law is expressed using $\mathbf{E}=-\nabla_\mathbf{x}\phi$ as a Poisson equation for the electric potential $\phi$ of the form
\begin{equation}
\nabla_\mathbf{x}^2 \phi = -\f{e(n_i-n_e)}{\varepsilon_0},
\end{equation}
where $\varepsilon_0$ and $e$ respectively denote the vacuum permittivity and elementary charge, and $n_i$ and $n_e$ respectively denote the ion and electron number densities
\begin{equation}
n_i(\mathbf{x};t) = \int_\mathbf{v} f_i(\mathbf{x},\mathbf{v'};t) \, \diff \mathbf{v'}; \qquad 
n_e(\mathbf{x};t) = \int_\mathbf{v} f_e(\mathbf{x},\mathbf{v'};t) \, \diff \mathbf{v'}.
\end{equation}
Periodic and no-flux boundary conditions are employed for $\mathbf{x}$ and $\mathbf{v}$, respectively.
\setcitestyle{notesep={, }}

\subsection{Problem setup and linear stability analysis}

We consider 1D1V and 2D2V current-carrying instabilities, where D and V denote the configuration and velocity spaces, respectively. The corresponding simulations are respectively two- and four-dimensional. Since these are long-wavelength instabilities, we choose domain lengths sufficiently larger than the Debye length, $\lambda_D = \sqrt{(\varepsilon_0 k_\text{B} T_e)/(n_e e^2)}$. The species temperatures are $T_i = 0.2\text{ eV}$ and $T_e = 2\text{ eV}$, the electrons have a net axial drift described by the initial electron Mach number $M_{e,d} = U_e/c_{e,d}$, and the ions have zero initial mean speed. Lengths and velocities are respectively nondimensionalized by $\lambda_D$ and the $s$-dimensional thermal speed, $c_{*,s} = \sqrt{s k_\text{B} T_*/m_*}$ (so $M_e = M_{e,1}$ and $c_* = c_{*,1}$), while time is nondimensionalized by the inverse electron frequency $1/\omega_e = \lambda_D/c_e$.

Linear growth rates for the current-carrying instability can be analytically predicted via solution of the (dimensional) linear dispersion relation for electrostatic waves
\begin{equation}
1 + \sum_* \f{\omega_*^2}{c_*^2k^2}\left[1 + \left(\f{\omega}{\sqrt{2}c_* k} - \f{\delta_{*e} M_{e,1}\cos\theta}{\sqrt{2}}\right) Z\left(\f{\omega}{\sqrt{2}c_* k}  - \f{\delta_{*e} M_{e,1}\cos\theta}{\sqrt{2}}\right)  \right],\label{eqn:disp}
\end{equation}
where $k = |\mathbf{k}|$ and $\omega$ are, respectively, the modal wavenumber magnitude and angular frequency, $\theta$ is the angle between $\mathbf{k}$ and the axial direction $x$, $\omega_* = \sqrt{(n_* e^2)/(m_* \varepsilon_0)}$ is the plasma frequency, $c_* = \sqrt{k_\text{B} T_*/m_*}$, and $Z$ is the plasma dispersion function
\begin{equation}
Z(\xi) = \f{1}{\sqrt{\pi}}\int_{-\infty}^{\infty} \f{e^{-z^2}}{z - \xi} \, \diff z = \ii \sqrt{\pi} e^{-\xi^2} \text{erfc}(-\ii \xi); \qquad \f{\diff Z(\xi)}{\diff \xi} = -2\left[1 + \xi Z(\xi)\right].
\end{equation}

\section{1D current-carrying instability}\label{sec:1D}

\begin{figure}
\begin{center}
    	\includegraphics[trim = 0 20 0 10, width = \textwidth]{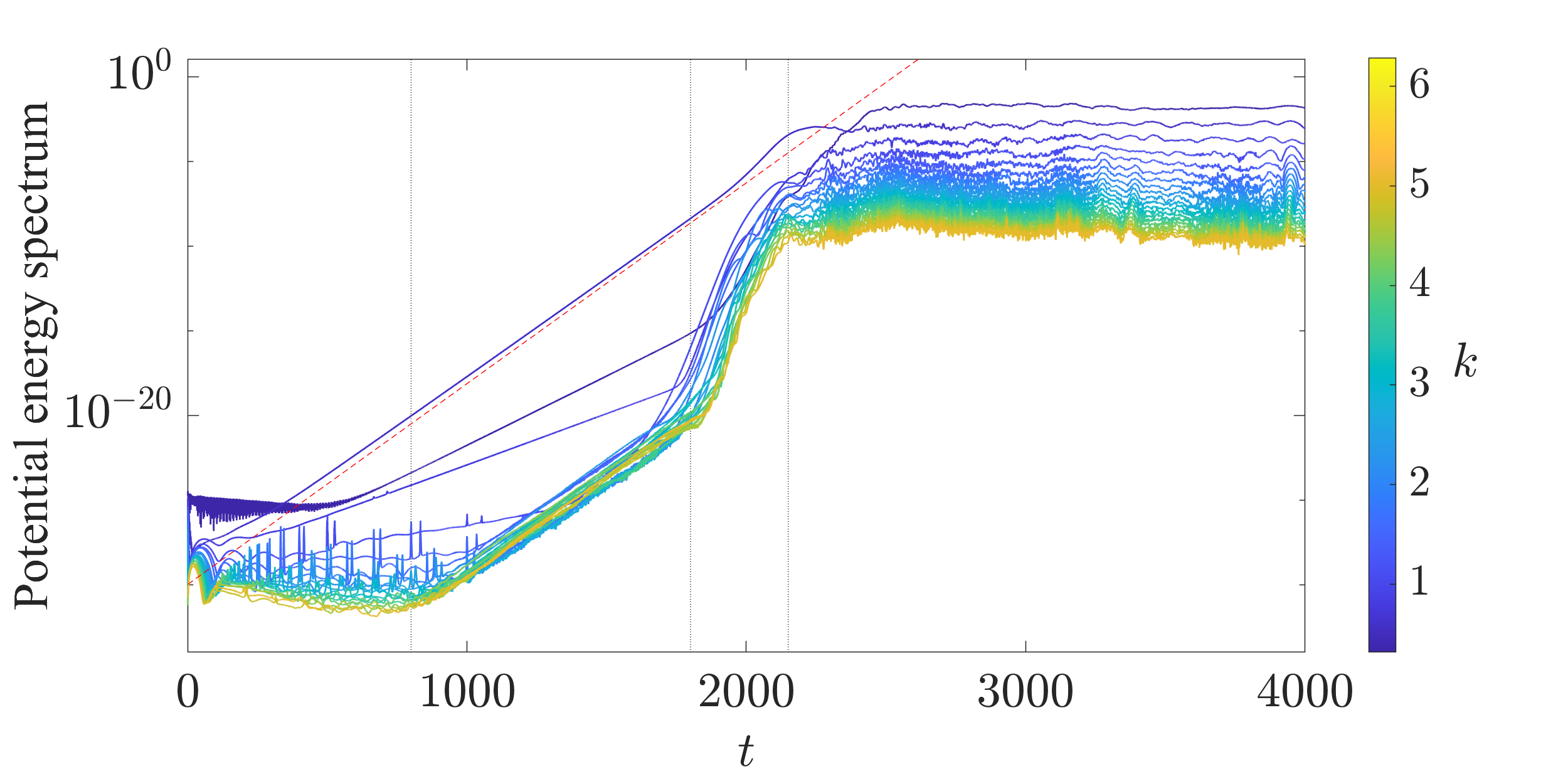}
      	\caption{Time evolution of the ensemble-averaged electrostatic potential energy spectrum. Energies are normalized by $\phi_\text{th}/\lambda_D$, where $\phi_\text{th}=k_\text{B} T_e/e$ is the thermal potential. Hereinafter, lengths and times are nondimensionalized by $\lambda_D$ and $1/\omega_e$, respectively. The sloped dashed line denotes the maximum growth rate obtained from Eq.~\eqref{eqn:disp}, while the vertical dotted lines qualitatively demarcate different evolution stages. Every fifth mode is plotted and the curves are colored from blue to yellow (dark to light in grayscale) in increasing $k$.\label{fig:spec1D}}
\end{center}
\end{figure}

\begin{figure}
\begin{center}
    	\includegraphics[trim = 0 30 0 5, width = \textwidth]{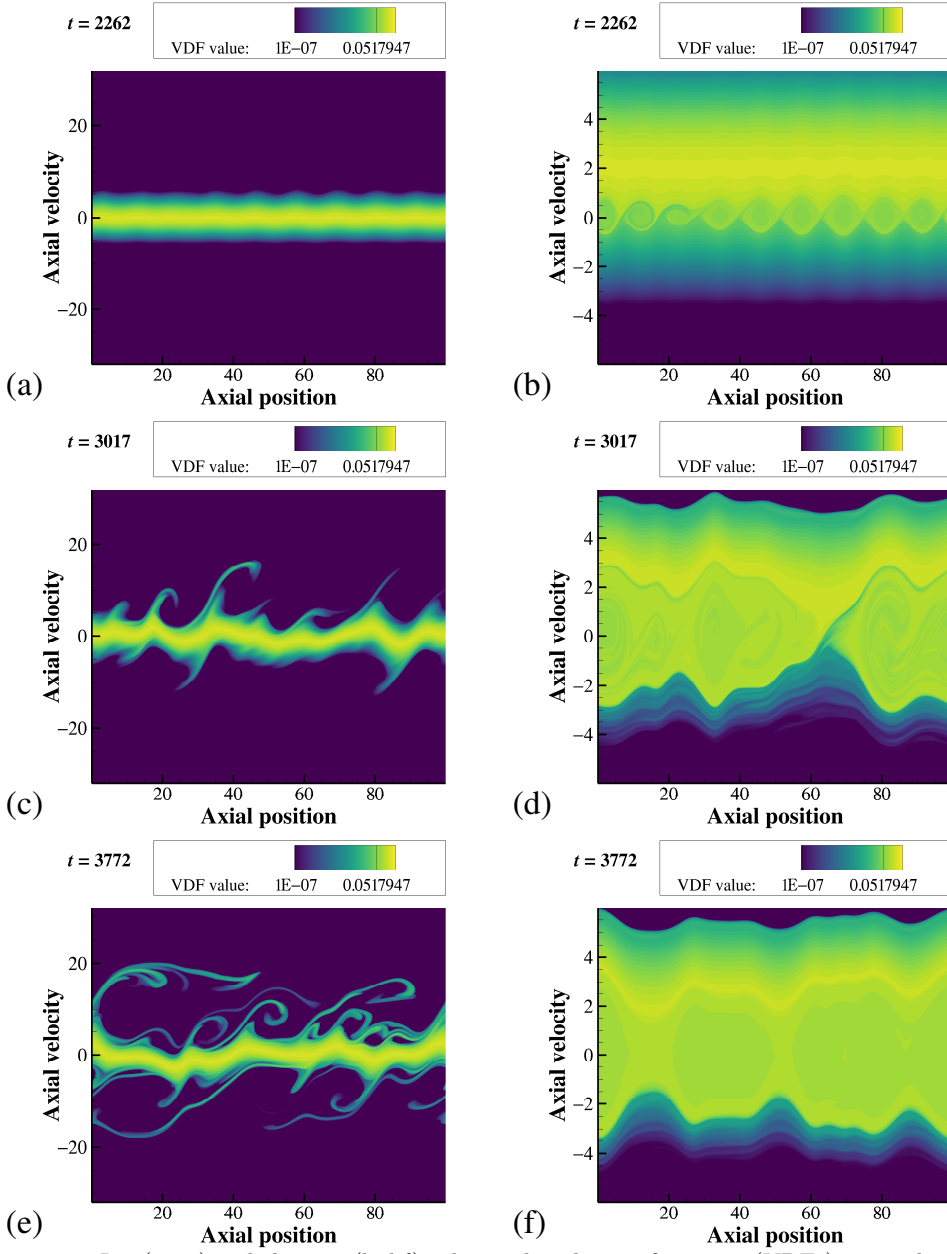}
      	\caption{Ion (a,c,e) and electron (b,d,f) velocity distribution functions (VDFs), normalized by the initial spatially uniform number density, at $t=2.3\times10^3$ (a,b), $3.0\times10^3$ (c,d), and $3.8\times10^3$ (e,f). Note that the contours are logarithmically spaced.\label{fig:vdf1D}}
\end{center}
\end{figure}

To interpret the 2D current-carrying instability more easily, we first discuss pertinent results of the 1D instability. Six 1D1V simulations with domain extents $x \in [0, 100], \, v_{x,i} \in [-32, 32], \, v_{x,e} \in [-6, 6]$ and resolutions $\Delta x = 1/10, \, \Delta v_{x,i} = 1/25, \, \Delta v_{x,e} = 1/100$ for the spatial, ion velocity, and electron velocity dimensions, respectively, were performed with $\Delta t = 0.016$ and $M_{e,1}=2.0$ in line with the grid-point recommendations of \citet{chan2022grid} for instability resolution. The time evolution of the ensemble-averaged electrostatic potential energy spectrum, obtained through a Fourier decomposition of $\phi$, is plotted in Figure~\ref{fig:spec1D}. Four stages of evolution are discernible. Modes first grow linearly at their analytically predicted modal growth rates. Harmonics then interact with the fastest-growing fundamental to grow at a comparable rate. The remaining modes eventually experience accelerated growth, filling in the intermediate wavenumbers to form a saturated and persistent broadband spectrum. The growth of harmonics (locking) and subsequent fill-in are features of turbulence also seen in, e.g., hydrodynamic simulations of turbulent boundary layers. Note that larger modes always lead and exceed smaller modes, possibly casting in doubt the existence of an inverse energy cascade.

Figure~\ref{fig:vdf1D} plots the ion and electron velocity distribution functions at three time instances: one in the third stage (nonlinear fill-in) and two in the fourth stage (saturated turbulence). Clockwise vortical motion represents trapping. The excitation of trapped electron oscillations precedes the generation of high-energy ions. Forward-streaming ions and electrons are sustained at the early stage of saturated turbulence, with backward-streaming ions and electrons following about $O(10)$ ion oscillation periods after.

\subsection{Interscale phase-space transfer and locality}

\begin{figure}
\begin{center}
    	\includegraphics[trim = 0 15 0 10, width = \textwidth]{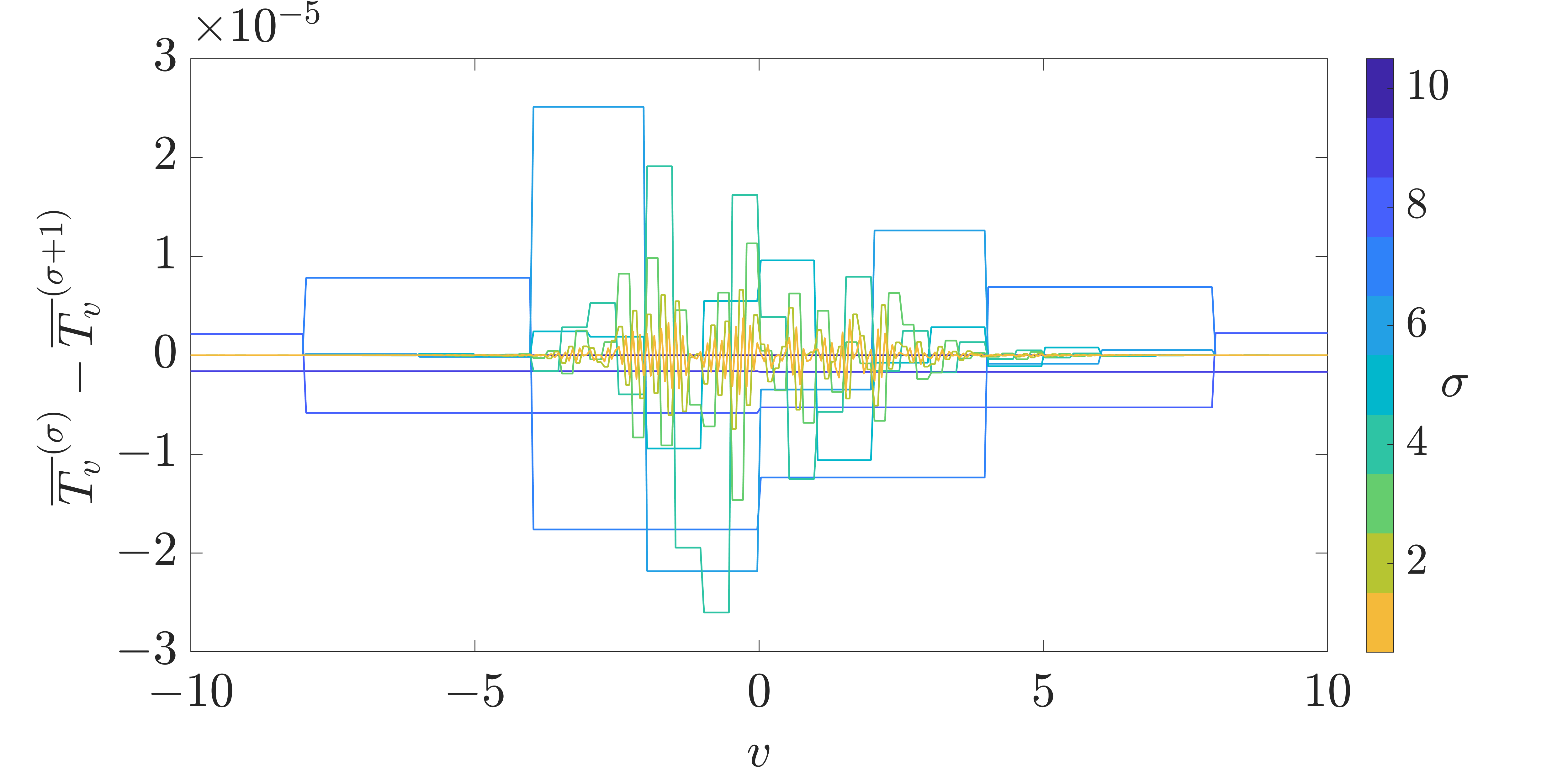}
      	\caption{The ensemble, time, and spatially averaged bandpass-filtered interscale transfer term due to velocity transport at different scales $\sigma$.\label{fig:Tv}}
\end{center}
\end{figure}

Preliminary analysis of interscale transfer of $f_i$ is performed for 1D1V phase space through analysis of the following transport equation for the filtered ion variance $\ov{f_i}^2/2$
\begin{multline}
\f{\partial}{\partial t}\left(\f{1}{2}\ov{f_i}^2\right) + \f{\partial}{\partial x}\left(\ov{v}\cdot\f{1}{2}\ov{f_i}^2\right) + \f{q_i}{m_i}\f{\partial}{\partial v}\left(\ov{E}\cdot\f{1}{2}\ov{f_i}^2\right) = -\f{\partial}{\partial x}\left[\ov{f_i}\left(\ov{vf_i}-\ov{v}\ov{f_i}\right)\right]-{}\\
{}-\f{q_i}{m_i}\f{\partial}{\partial v}\left[\ov{f_i}\left(\ov{Ef_i}-\ov{E}\,\ov{f_i}\right)\right] + \left(\ov{vf_i}-\ov{v}\ov{f_i}\right)\f{\partial \ov{f_i}}{\partial x} + \f{q_i}{m_i}\left(\ov{Ef_i}-\ov{E}\,\ov{f_i}\right)\f{\partial \ov{f_i}}{\partial v}.
\end{multline}
The last two terms are denoted $\ov{T_x}$ and $\ov{T_v}$, and represent interscale transfer due to transport in the physical and velocity spaces, respectively. The overbar denotes a filtering operation, which is performed here with the assistance of a discrete wavelet decomposition. The ion distribution function $f_i$ may be written as~\citep[cf.][]{kim2018spatially}
\begin{equation}
f_i(x,v) = \sum_{s=1}^{\mathcal{S}}\sum_{d=1}^{3}\sum_{x_s,v_s} \check{f}^{(s,d)}(x_s,v_s)\mathcal{G}^{(s,d)}(x-x_s,v-v_s) + \sum_{x_\mathcal{S},v_\mathcal{S}}\hat{f}^{(\mathcal{S})}\mathcal{H}^{(S)}(x-x_s,v-v_s),
\end{equation}
where $s$ and $\mathcal{S}$ respectively denote the scale index and number of scales, $d$ is a wavelet directionality index, $\mathcal{G}^{(s,d)}$ and $\mathcal{H}^{(s)}$ respectively denote the wavelet and scaling functions, and $\check{f}^{(s,d)}$ and $\hat{f}^{(s)}$ respectively denote the detail and approximation coefficients. Here, the Haar wavelet is used and the Vlasov simulation is performed with $1{,}024$ points in each dimension so that $\mathcal{S}=10$. Then, $\ov{f_i}^{(\sigma)}$ is defined for $\sigma = 1,\ldots,\mathcal{S}$ as
\begin{equation}
\ov{f_i}^{(\sigma)}(x,v) = \sum_{s=\sigma}^{\mathcal{S}}\sum_{d=1}^{3}\sum_{x_s,v_s} \check{f}^{(s,d)}(x_s,v_s)\mathcal{G}^{(s,d)}(x-x_s,v-v_s) + \sum_{x_\mathcal{S},v_\mathcal{S}}\hat{f}^{(\mathcal{S})}\mathcal{H}^{(S)}(x-x_s,v-v_s).
\end{equation}
Figure~\ref{fig:Tv} plots $\ov{T_v}^{(\sigma)}-\ov{T_v}^{(\sigma+1)}(v)$, averaged over all $x$, late $t \in [3.8\times10^3,7.6\times10^3]$, and 40 ensemble realizations. This represents the ion variance gained or lost at velocity $v$ and scale $\sigma$ due to transfer between velocities and scales. The corresponding spatial interscale term is negligible in comparison (not shown here). At these late times, interscale transfer is biased towards negative velocities, as backward-streaming ions are formed on average after forward-streaming ions. The limited correlation between large and small scales hints at interscale locality, which may be verified through an information-theoretic approach analyzing causality~\citep{lozanoduran2022information}.

\section{2D current-carrying instability}\label{sec:2D}

\begin{figure}
\begin{center}
    	\includegraphics[trim = 0 165 0 145, width = \textwidth]{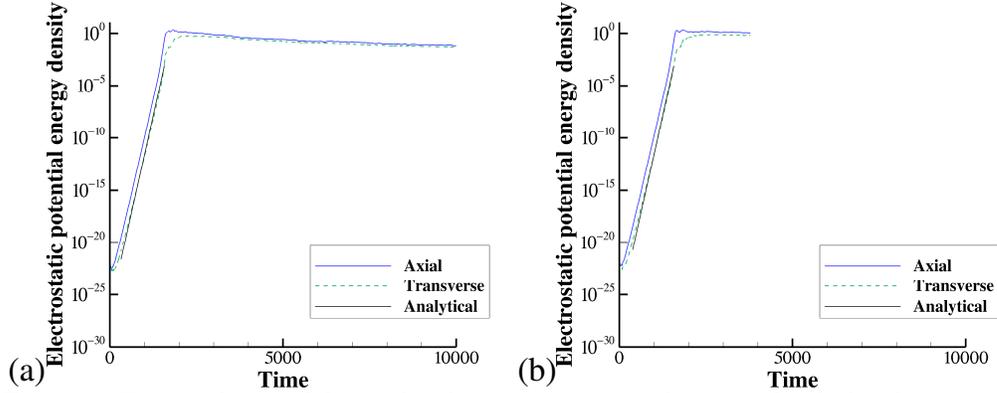}
      	\caption{Time evolution of the axial and transverse potential energies for the baseline (a) and velocity-refined (b) simulations. The analytical line denotes the growth rate from Eq.~\eqref{eqn:disp}. \label{fig:epe2D}}
\end{center}
\end{figure}

Building on the preliminary work of~\citet{vazsonyi2021deterministic}, the 2D2V instability is simulated with domain extents $x,y \in [0, 80], \, v_{x,i}, v_{y,i} \in [-32, 32], \, v_{x,e}, v_{y,e} \in [-6, 6]$ and resolutions $\Delta x = \Delta y = 1/2.5, \, \Delta v_{x,i} = \Delta v_{y,i} = 1/2, \, \Delta v_{x,e} = \Delta v_{y,e} = 1/16$ for the spatial, ion velocity, and electron velocity dimensions, respectively. A second simulation was performed with twice the resolution in all velocity dimensions to ascertain velocity grid convergence, given that resolutions are decreased from the 1D case for computational tractability. The simulation is converged with respect to the spatial grid and domain extent for the quantities of interest (not shown here). Both simulations were performed with $\Delta t = 0.062$ and $M_{e,2}=1.6$ (so $M_{e,1}=2.3$). Figure~\ref{fig:epe2D} plots the time evolution of the axial and transverse potential energies, respectively, $\sum E_x^2/2$ and $\sum E_y^2/2$, where $E_x$ and $E_y$ are the axial and transverse electric field strengths in each cell. The baseline resolution is seen to exhibit grid convergence and is analyzed for the remainder of this work.

\begin{figure}
\begin{center}
    	\includegraphics[trim = 40 0 40 0, width = 0.9\textwidth]{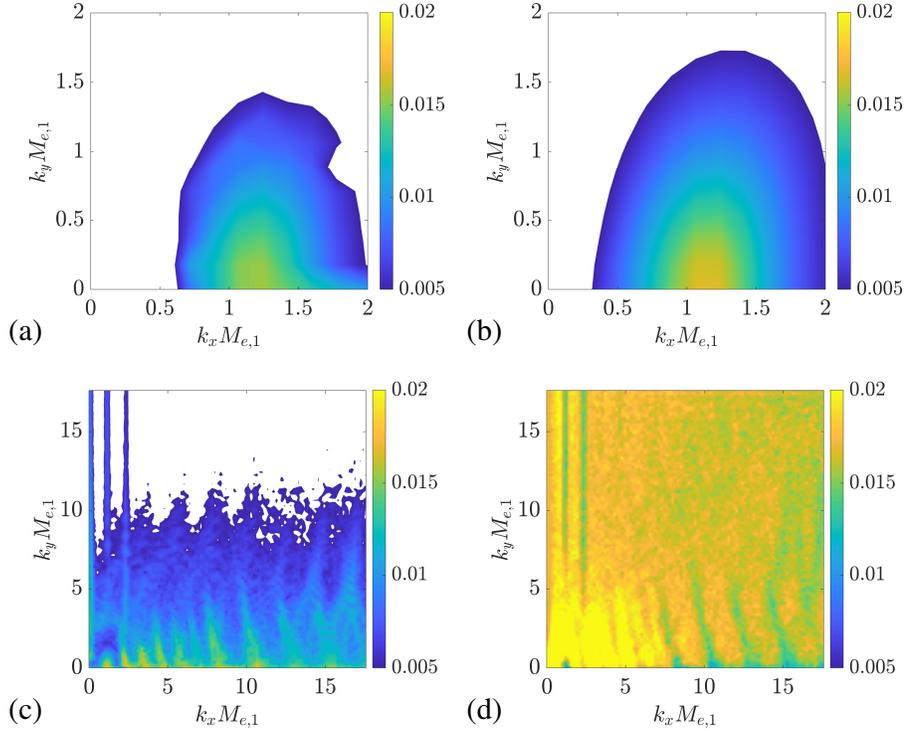}
      	\caption{Numerical growth rates of spectral modes $\{k_x M_{e,1},k_y M_{e,1}\}$ of the electric potential spectrum $E_{\phi\phi}$ for $t=[0,3.8\times10^2]$ (a), $[3.8\times10^2,1.1\times10^3]$ (c), and $[1.1\times10^3,1.9\times10^3]$ (d), obtained via linear regression. The maximum growth rate predicted by Eq.~\eqref{eqn:disp} is 0.017, and the corresponding analytical modal growth rates are plotted in (b). Growth rates less than $5\times10^{-3}$ are excluded to remove cases where oscillations confound the regression.\label{fig:growth2D}}
\end{center}
\end{figure}

Figure~\ref{fig:growth2D} plots the modal growth rates in time intervals qualitatively corresponding to the four stages identified in Figure~\ref{fig:spec1D}. The wavenumbers are multiplied by $M_{e,1}$ for direct comparison with~\citet{amano2009nonlinear}, particularly the linear stage in Figure~\ref{fig:growth2D}(a,b). The same qualitative trends are observed in 1D and 2D: linear growth as described by Eq.~\eqref{eqn:disp}, subsequent comparable growth of harmonics, nonlinear fill-in of intermediate wavenumbers via a catch-up mechanism, and eventually saturated turbulence.

\begin{figure}
\begin{center}
    	\includegraphics[trim = 60 10 70 5, width = \textwidth]{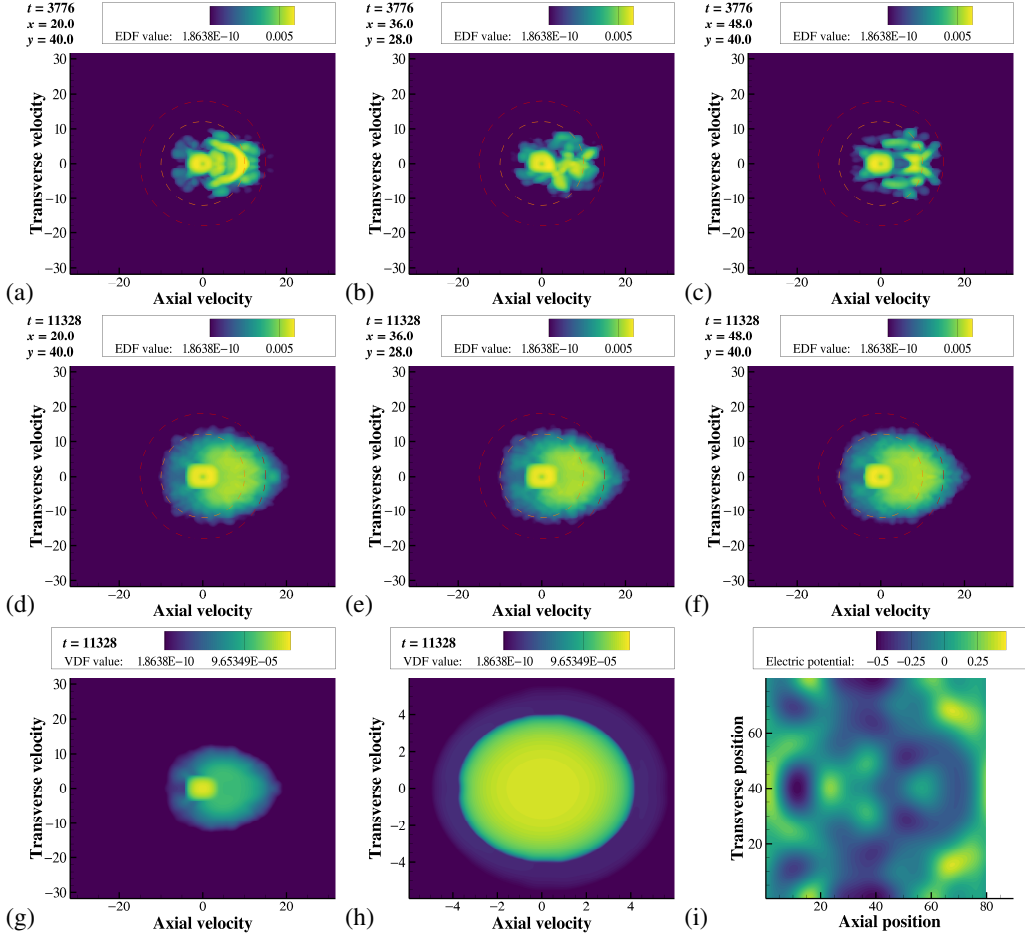}
      	\caption{Local ion energy distribution functions (EDFs) at three different locations at $t = 3.8\times10^3$ (a,b,c) and $t=1.1\times10^4$ (d,e,f). The spatially averaged ion (g) and electron (h) velocity distribution functions (VDFs), as well as the electric potential profile (i), are also plotted for $t=1.1\times10^4$. The contours in (a--h) are logarithmically spaced, and the two concentric circles in (a--f) represent 20 and 45 eV contours.\label{fig:vdf2D}}
\end{center}
\end{figure}

Figure~\ref{fig:vdf2D} plots several representative ion and electron velocity and energy distribution functions. Electron trapping and isotropization occur more rapidly than their corresponding ion processes owing to the larger thermal speed and smaller response time of electrons. High-energy ions are generated in abundance at equivalent temperatures of between 20 and 50 eV with practical relevance to hollow cathode sputtering.

\section{Conclusions}\label{sec:conc}

The investigation of current-driven plasma instabilities is crucial to determine the origin and fluxes of high-energy ions that cause hollow cathode erosion in electric spacecraft thrusters. A direct kinetic solver is used to study the evolution of the ion and electron velocity distribution functions without contamination from statistical noise inherent in state-of-the-art particle methods. Both 1D and 2D current-driven instabilities exhibit four developmental stages: linear growth, quasilinear resonance, nonlinear fill-in, and saturated turbulence. The maximum linear modal growth rate matches analytical predictions from the linear plasma dispersion relation. Harmonics of the fastest-growing fundamental, followed by intermediate wavenumbers, grow in a process that resembles the development of hydrodynamic turbulence. 2D instabilities further exhibit a return to isotropy also reminiscent of classical fluid behavior. However, unlike hydrodynamic turbulence, which only fully emerges in 3D, such plasma turbulence occurs even in 1D and 2D instabilities as postulated by~\citet{buneman1959dissipation} since ions and electrons are allowed to interpenetrate unlike fluids. While the potential energy quickly saturates in the turbulent regime, backward-streaming ions are only formed after several ion trapping cycles. More generally, direct kinetic solvers can be used to provide quantitative predictions of cathode sputtering rates and potential fluctuations observable in experiments.

\subsection*{Acknowledgments}
The authors would like to acknowledge K. Griffin, K. Schneider, K. Matsuda, and the multiphase group at the CTR Summer Program for helpful discussions. This work utilized the Blanca condo computing resource of the University of Colorado, as well as the Summit supercomputer, which is supported by the National Science Foundation (awards ACI-1532235 and ACI-1532236), the University of Colorado, and Colorado State University.


\end{document}